\documentclass{aa}
\usepackage{graphicx}
\usepackage{natbib}
\usepackage{amsmath}
\bibpunct{(}{)}{;}{a}{}{,}

\newcommand{\xb}{\boldsymbol{x}}

\newcommand{\bb}{\boldsymbol{b}}
\newcommand{\Ob}{\boldsymbol{0}}
\newcommand{\qb}{\boldsymbol{q}}
\newcommand{\db}{\boldsymbol{d}}
\newcommand{\Dc}{\mathcal{D}}
\newcommand{\vtw}{{\it VTW}}
\newcommand{\shm}{{\it SHM}}
\newcommand{\bshm}{{\it BSHM}}

\begin{document}

\title{Some Good Reasons to Use Matched Filters \\ for the Detection of Point Sources in CMB Maps}

   \author{R. Vio\inst{1}
          \and
          P. Andreani\inst{2}
          \and
          W. Wamsteker\inst{3}
          }

   \offprints{R. Vio}

   \institute{Chip Computers Consulting s.r.l., Viale Don L.~Sturzo 82,
              S.Liberale di Marcon, 30020 Venice, Italy\\
              ESA-VILSPA, Apartado 50727, 28080 Madrid, Spain\\
              \email{robertovio@tin.it}
         \and
              Osservatorio Astronomico di Padova, vicolo dell'Osservatorio 5,
                  35122 Padua, Italy \\
              \email{andreani@pd.astro.it}
         \and
             ESA-VILSPA, Apartado 50727, 28080 Madrid, Spain\\
             \email{willem.wamsteker@esa.int}
             }

\date{Received .............; accepted ................}

\abstract{In this paper we comment on the results concerning the performances of matched filters, scale
adaptive filters and Mexican hat wavelet that recently appeared in literature in the context of
point source detection in Cosmic Microwave Background maps. In particular, we show that,
contrary to what has been claimed, the use of the matched filters still appear to be the most reliable
and efficient method to disantangle point sources from the backgrounds, even when using
detection criterion that, differently from the classic $n\sigma$ thresholding rule, takes into account
not only the height of the peaks in the signal corresponding to the candidate sources but also their curvature.
}
\titlerunning{Optimal Detection of Sources}
\authorrunning{R. Vio, P. Andreani, \& W. Wamsteker}
\maketitle

\section{Introduction}

Studying diffuse backgrounds in all-sky maps implies the possibility of disentangling background signals
from those originated from point sources. This task is of fundamental importance in
dealing with Cosmic Microwave Background (CMB) data. 
In this context various papers studied the``optimal'' method for such a task. Three main methods have been
considered so far: the Mexican hat wavelet (Cayon et al., 2000), the scale-adaptive filters (or optimal pseudo-filters)
and the matched filters (Sanz et al., 2001, Vio et al., 2002 and reference therein).
Matched filter (MF) is constructed taking into account the source profile and the background
to get the maximum signal-to-noise ratio (SNR) at the source position. Scale-adaptive filter (SAF)
is built similarly to MF with the additional constraint to have 
a maximum in filtered space at the scale and source position.
The Mexican hat wavelet (MHW) represents a separate
case since it is ``a priori'' filter, adapted to the detection of point sources. Its main limitation
is that it is founded on semi-empirical arguments and therefore lacks a rigorous theoretical justification. 
For this reason, in the following we will be especially concerned with MF and SAF.

Vio et al. (2002) (henceforth {\vtw}) have shown that, in spite the claims of ``optimality'' for SAF and MHW
(Sanz et al., 2001, henceforth {\shm}), in reality these filters do not behave as good as the MF.
In a recent work in the context of one-dimensional signals, 
Barreiro et al. (2003, henceforth {\bshm}) compare SAF, MHW, and MF on the basis of a detection
criterion based on the Neyman-Pearson 
decision rule, that takes into account not only the height of signal peaks but also their curvature. 
These authors find that, although MF is effectively optimal in most of the
cases, there are situations where SAF and MHW can overperform it. Here we show that such a result 
is not correct since it is linked to the measure of performance adopted by authors, that tends to favour the 
filters characterized by a low detection capability. MF is in general superior to these other two filters. 

\section{Problem Formalization} \label{sec:formalization}

For sake of generality, we firstly present our arguments in $R^n$ and then we specialize the results to 
the one-dimensional case. 

The sources are assumed to be point-like signals convolved with the beam of the measuring instrument and are
thus assumed to have a profile $\tau(\xb)$. The signal $y(\xb)$, $\xb\in R^n$, is modeled as
\begin{equation}\label{datamodel}
y(\xb) = \sum_{j} s_j(\xb) + z(\xb)
\end{equation}
where
\begin{equation}
s_j(\xb)=A_j\, \tau(\xb -\xb_j),
\end{equation}
$A_j$ and $\xb_j$ are, respectively, unknown source amplitudes and locations, and $z(\xb)$ is a zero-mean background
with power-spectrum $P(\qb)$
\begin{equation} \label{eq:powerz}
{\rm E}\,[\,z(\qb)\, z^*(\qb')\,] = P(\qb) ~\delta^n(\qb - \qb').
\end{equation}
Henceforth ${\rm E}[\cdot]$ and ``$~{}^*~$'' will denote the expectation and complex conjugate operators, respectively,
$\delta^n(\qb - \qb')$ the $n$-dimensional Dirac distribution, and $z(\qb)$ the Fourier transform of $z(\xb)$
\begin{equation}
z(\qb) = \int_{-\infty}^{+\infty} z(\xb) ~{\rm e}^{- i \qb \cdot \xb} ~\db\xb.
\end{equation}
To properly remove the point sources from the signal it is necessary to estimate the locations  $\{\xb_j\}$ and
amplitudes (fluxes) $\{A_j\}$ of the sources.

The classic procedure for the detection of the sources consists in filtering signal to enhance 
the sources with respect to the background. This is done by cross-correlating the signal $y(\xb)$ 
with a filter $\Phi$. The source locations are then determined by selecting the peaks in the filtered signal
that are above a chosen threshold. Finally, the source amplitudes are estimated as the
values of the filtered signal at the estimated locations. The question is the selection of an
optimal filter $\Phi$ for such procedure. In order to define it, some assumptions are necessary.
In particular it is assumed that the source profile and background spectrum are known,
the profile is spherically symmetric, characterized by a scale $R_s$, and the background is
isotropic. These assumptions allow to write
$s(\xb) \equiv s(x)$, where $x = \| \xb \|$, and $P(\qb) \equiv P(q)$ for $q = \| \qb \|$. In addition,
source overlap is assumed negligible. In the present context, we are interested in
the general family of spherically symmetric filters $\Phi(\xb; \bb)$ 
of the form $\Phi(\xb;\bb) = \phi(\, \| \xb - \bb \|\, )$.
The cross-correlation between $\Phi(\xb; \bb)$ and $y(\xb)$ provides a filtered
field $w(\bb;\phi)$ with mean $\mu(\bb;\phi)$ and variance  $\sigma^2(\phi)$.

\subsection{Matched filters (MF)} \label{sec:MF}

Source locations are assumed to be known
and the aim is to estimate the amplitudes. Given the assumed distance between the sources, it is enough to
consider
a field $y(\xb)$ as in (\ref{datamodel}) with a single source at the origin, $s(\xb) = A\,\tau(\xb)$.
Its amplitude $w(\Ob;\phi)$ is estimated by requiring it to be an unbiased estimator of $A$, i.e.,
$\mu(\Ob;\phi) = A$. On the other hand, to enhance the magnitude of the source relative to the background  
the filter $\Phi$ is required to minimize
the variance $\sigma^2(\phi)$. This has the effect of maximizing, among unbiased estimators,
the detection level
\begin{equation} \label{eq:detection}
\mathcal{D}(\phi)=\frac{\mu(\Ob;\phi )}{\sigma(\phi)},
\end{equation}
which measures the capability of the filter to  detect correctly a source at the prescribed location.

Since $\Phi$ is chosen in a way that $w(\Ob;\phi)$ is a minimum variance linear (in $y(\xb)$) unbiased
estimator of $A$, it follows that (Gauss-Markov theorem) $w(\Ob;\phi)$ is the (generalized) least squares estimate
of $A$ achieved by the filter
\begin{equation} \label{eq:filter2}
\phi(q) = \frac{1}{\delta a} ~ \frac{\tau(q)}{P(q)},\qquad a \equiv \int_0^{+\infty}q^{n-1} \frac{\tau^2}{P} ~dq,
\end{equation}
with minimum variance
\begin{equation} \label{eq:sigma2}
\sigma^2(\phi) = \frac{1}{\delta a},
\end{equation}
where $\delta = 2\, \pi^{n/2}\,\Gamma^{-1}(n/2)$.
In other words, filter (\ref{eq:filter2}), called {\it matched filter}, optimizes the signal-to-noise ratio
\citep[e.g.,][]{kozma,pratt91}. Although this filter is commonly used for signal
detection, it is not the only approach to define an
"optimal filter". A possible alternative is represented by the Wiener-filters that, however,
are designed to minimize the prediction error given covariance information \citep{rab75}.

\subsection{Scale adaptive filters (SAF)}
\label{sec:SAF}
In the pseudo-filter approach of {\shm} the filters have the same form
of $\Phi(\xb; \bb)$ with an additional scale
dependence
\begin{equation} \label{eq:psi}
\Psi(\xb; R, \bb) = \frac{1}{R^N}\, \psi \left( \frac{\| \xb - \bb \|}{R} \right),
\end{equation}
for some spherically symmetric function $\psi$.
The cross-correlation between this filter at scale $R$ and $y(\xb)$ provide a filtered field $w(\bb,R;\psi)$.

To determine an optimal filter $\psi$, {\shm}  minimize the variance of the filtered field
with the two constraints: $w(\Ob,R_0;\psi)$ is required to be, as in the previous section,
an unbiased estimator of $A$ for some known $R_0 \approx R_s$, and $\psi$ is selected so
that $\mu(\Ob,R;\psi)$
has a local maximum at scale $R_0$. This latter translates into
\begin{equation} \label{eq:constr2}
\int_0^{+\infty} q^{n-1} \tau(q) \,\psi(R_0 q) \left(n + \frac{d \ln \tau}{d \ln q} \right)~dq = 0.
\end{equation}
Minimizing $\sigma^2(R_0;\psi)$ with the two constraints yields the filter ({\shm})
\begin{equation} \label{eq:psi1}
\psi(R_0 q) = \frac{1}{\delta ~\Delta} ~\frac{\tau(q)}{P(q)} \left[ nb + c -(na+b)\,
\frac{d \ln \tau(q)}{d \ln q} \right],
\end{equation}
where $\Delta = ac - b^2$,
\begin{equation}
\begin{aligned}
b &\equiv \int_0^{+\infty} q^{n-1} \frac{\tau}{P} ~\frac{d\tau}{d\ln q}, \\
c &\equiv \int_0^{+\infty} q^{n-1} \frac{1}{P} \left( \frac{d \tau}{d \ln q} \right)^2 ~dq,
\end{aligned}
\end{equation}
and $a$ is as in (\ref{eq:filter2}). This filter provides a field of variance
\begin{equation} \label{eq:sigma1}
\sigma^2(R_0;\psi) = \frac{n^2 a + 2 n b + c}{\delta \Delta},
\end{equation}
and an estimator of the amplitude $A$ that is again linear and unbiased.

\section{Filter comparison} \label{sec:comparison}

In their work {\vtw} stress the fact that, since both $\Phi$ and $\Psi$ provide a
linear and unbiased estimate of
the amplitude $A$ then, regardless the source profile and background spectrum
and because of the optimality of the least squares, $\sigma^2(R_0;\psi)\geq \sigma^2(\phi)$.
As a consequence the value of the detection level, $\Dc$, corresponding
to $\Phi$ is at least as high, or higher, than that achieved with $\Psi$.
Furthermore, via an extensive 
set of numerical simulation {\vtw} have shown that this conclusion holds even when the source
location uncertainty is taken into account. In other words, enough information
about the scale of the source is already included in the derivation of the matched filter.
Via numerical simulations {\vtw} have also shown that MF overperforms SAF when comparing the resulting
numbers of incorrectly detected sources. {\vtw}'s conclusion is then nothing is gained 
by using SAF.

Recently, in the context of one-dimensional signals, zero-mean Gaussian background with scale-free power spectrum 
$P(q) = D q^{-\gamma}$, and Gaussian profile $\tau(x) = A {\rm e}^{- x^2 / 2 R_0^2}$ for the source, {\bshm} 
criticized
these conclusions through the argument that the detection level $\mathcal{D}$ and the
$n\sigma$ thresholding method used by {\vtw} as detection rule are not sufficient to support 
their results \footnote{Here it is necessary to stress that, contrary to what written by {\bshm}, the 
superiority of MF with respect to SAF is claimed not only on the basis of the larger value of $\mathcal{D}$
and of the number of correct detections, but also on the better detection capability for a given average number 
of incorrect detections.}. For this reason, they introduce a new detection criterion based on a Neyman-Pearson 
decision rule which uses not only the heigth of the maxima in the signal but also their curvature. This
method can be summarized as follows (for more details, see {\bshm})

\subsection{First case: fixed source amplitudes}

If the $1D$ background $z(x)$ is Gaussian, then it is possible to estimate the expected total number density $n_b$ 
of maxima (i.e., number of maxima per unit interval in $x$) as well their expected number density $n_b(\nu, \kappa)$ 
per intervals $(\nu, \nu+d\nu)$ and $(\kappa, \kappa+d\kappa)$,
where $\nu \equiv z/\sigma_0$ and $\kappa \equiv -z''/\sigma_2$ are the normalized field and
curvature, respectively. Here, $\sigma_n^2$ is the moment of order $2n$ associated with the field.
If all the sources are assumed to have the same amplitude $A$,
it is possible to estimate the corresponding quantities $n$ and $n(\nu, \kappa | \nu_s)$, $\nu_s = A/\sigma_0$,
when the sources are embedded in the background. These quantities
allow to calculate, for any region $R_*(\nu, \kappa)$, the probability density functions
\begin{equation}
p_b(\nu, \kappa | 0) = \frac{n_b(\nu, \kappa)}{n_b}, \quad p(\nu, \kappa | \nu_s) = \frac{n(\nu, \kappa)}{n},
\end{equation}
that can be interpreted as the probability that a given maximum is due to the background or to a local source,
respectively. In their turn, $p_b(\nu, \kappa | 0)$ and $p(\nu, \kappa | \nu_s)$ allow the calculation of
the quantities
\begin{equation} \label{eq:alpha}
\alpha = \int_{R_*} p_b(\nu, \kappa | 0) d\nu d\kappa
\end{equation}
\begin{equation} \label{eq:beta}
\quad 1-\beta = \int_{R_*} p(\nu, \kappa | \nu_s) d\nu d\kappa,
\end{equation}
that provide the so called {\it false alarm probability} (i.e., the 
probability of interpreting noise as signal) and the {\it power of the detection} (i.e., $\beta$
represents the probability of interpreting signal as noise). $R_*$ is called the {\it acceptance region}.

In order to obtain a detection criterion, {\bshm} introduce the significance $s^2$
\begin{equation} 
s^2 \equiv \frac{\left[ \langle N \rangle_{{\rm signal}} - \langle N \rangle_{{\rm no-signal}} \right]^2}
{\sigma_{{\rm signal}}^2 + \sigma_{{\rm no-signal}}^2},
\end{equation}
where in the numerator appears the difference between the mean number of peaks
in $N$ different realizations of the background, in presence and absence
of signal: $(1-\beta) N$ and $\alpha N$, respectively. $\sigma_{{\rm signal}}^2$ and 
$\sigma_{{\rm no-signal}}^2$ represent the variances of corresponding quantity $\langle N \rangle$.
It is not difficult to show that
\begin{equation}
s^2(\alpha,\beta) \propto \frac{(1 - \beta - \alpha)^2}{\beta (1-\beta) + \alpha(1-\alpha)}.
\end{equation}
The idea of {\bshm} is to maximize $s^2$ with respect to $\alpha$ with the constraint that $R_*$
is defined by the the Neyman-Pearson decision rule. The reason is that the acceptance region
\begin{equation} \label{eq:rule}
L(\nu, \kappa | \nu_s) \equiv \frac{p(\nu, \kappa | \nu_s)}{p(\nu, \kappa | 0)} \ge L_*
\end{equation}
provides the highest power $1-\beta$ for a given confidence level $\alpha$.
$L_*$ is a constant: if $L \ge L_*$ a source is present, whereas if $L < L_*$ no source is present. Eqs.~(\ref{eq:alpha})
and (\ref{eq:rule}) allow to exchange the maximization of $s^2$ with respect to $\alpha$ with the maximization
with respect to $L_*$.

\begin{figure}
        \resizebox{\hsize}{!}{\includegraphics{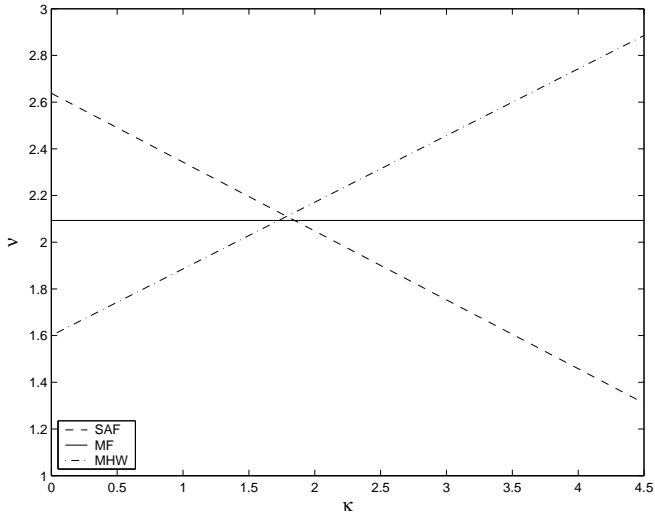}}
        \caption{Acceptance region $R_*$, when $\gamma=1.4$, for SAF, MF, and MHW for sources with aplitude
        $A$ such that $\nu_s=3$ after filtering with SAF. The maxima with $\nu$ and $\kappa$
        above the corresponding line are accepted as sources and those below are rejected.}
        \label{fig:criterion}
\end{figure}

It happens that for SAF, MF, and MHW, and independently from the index $\gamma$, $s^2$ is maximized for $L_* \approx 1$. 
Fig.~\ref{fig:criterion} shows the corresponding $R_*$ for sources with an amplitude A such as $\nu_s =3$ after filtering 
with SAF. This figure shows that, at variance with SAF and MHW, the acceptance region of MF
does not depend on the curvature $\kappa$ but only on the height of the maxima. Therefore, for MF the detection rule
proposed by {\bshm} provides a criterion similar to the classic $n\sigma$ thresholding rule. 

\begin{figure}
        \resizebox{\hsize}{!}{\includegraphics{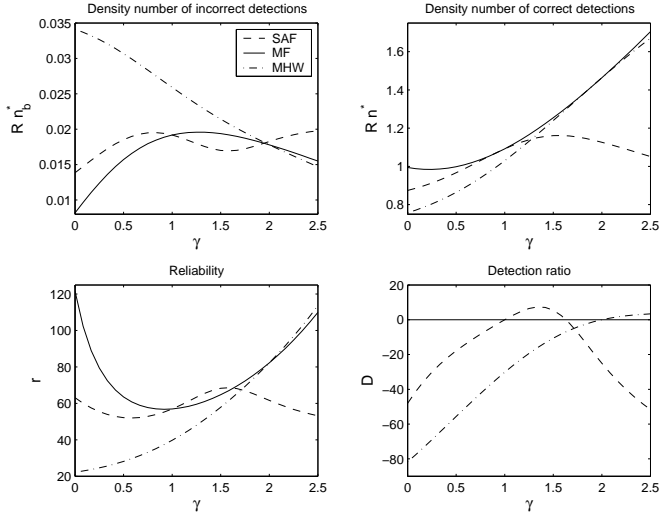}}
        \caption{Relationship $\gamma$ vs.  the number density $n_b^*$ of incorrect detections, the
        number density $n^*$ of correct detections, the
        {\it reliability} $r$, and the relative detection ratio $D$ corresponding to SAF, MF, and MHW for $L_*=1$.}
        \label{fig:compar1}
\end{figure}

Once fixed $R_*$ it is possible to calculate the expected number density $n_b^*$ of incorrect and the expected
number density $n^*$ of correct detections 
by integrating $n_b^*(\nu, \kappa)$ and $n(\nu, \kappa | \nu_s)$ over $R_*$. These quantities are used by {\bshm}
to calculate the ratio $r=n^*/n_b$, called {\it reliability}, and the quantity
\begin{equation}
D = \frac{r_i - r_{\rm MF}}{r_{\rm MF}} \times 100,
\end{equation} 
where subindex $i$ refers to the different filters, that these authors use as a measure of
the performances of MF, SAF, and MHW. Fig.~\ref{fig:compar1} shows $n_b^*$, $n^*$, $r$ and $D$, as
function of the index $\gamma$. 
Essentially on the basis of quantity $r$ {\bshm} claim that for 
$\gamma < 1$ and $\gamma > 1.6$ MF overperforms SAF, whereas
the contrary holds for $1 \le \gamma \le 1.6$. These conclusions deserve some comments. 

First, similarly to {\vtw} and in spite of the introduction of the 
new detection criterion, {\bshm} find that in most situations
the use of the second constraint (\ref{eq:constr2}) in SAF is not only useless but even harmful. 
Second, if, on the one hand, the superiority of MF for $\gamma < 1$ and $\gamma > 1.6$ 
is out of discussion (this filter provides the largest number of correct detections and the smallest number 
of incorrect ones),
the same conclusion for SAF when $1 \le \gamma \le 1.6$ is questionable. In this range of
$\gamma$ SAF provides a smaller number of incorrect detections, but at the same time also a 
smaller number of correct ones.
In this respect, at least in principle, the reliability parameter $r$ should be 
used as a measure of the filter performances only when an incorrect detection has a larger ``cost'' than
missing a source, a fact that has to be proved in the context of CMB. Furthermore, even in the case of an 
high ``cost'' for the incorrect detections, $r$ has to be used with great care.
The reason is that MF is constructed in such a way to maximize source detections. Therefore, the maximization
of $s^2$ with respect to $L_*$ provides a criterion favouring the detection of a true source rather than the 
rejection of a false one. 

\begin{figure}
        \resizebox{\hsize}{!}{\includegraphics{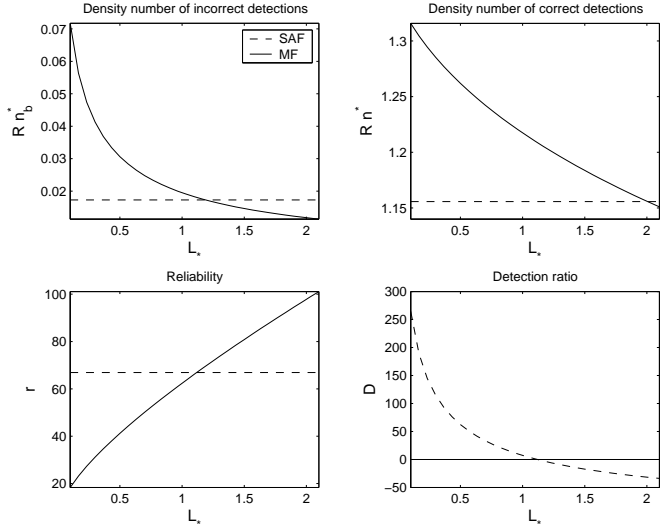}}
        \caption{Relationship $L^*$ vs. the number density $n_b^*$ of incorrect detections, the
        number density $n^*$ of correct detections, the
        {\it reliability} $r$, and the relative detection ratio $D$ corresponding to MF for $\gamma=1.4$. 
	  For comparison, the corresponding levels of SAF for $L_*=1$ are plotted too.}
        \label{fig:compar2}
\end{figure}

\begin{figure}
        \resizebox{\hsize}{!}{\includegraphics{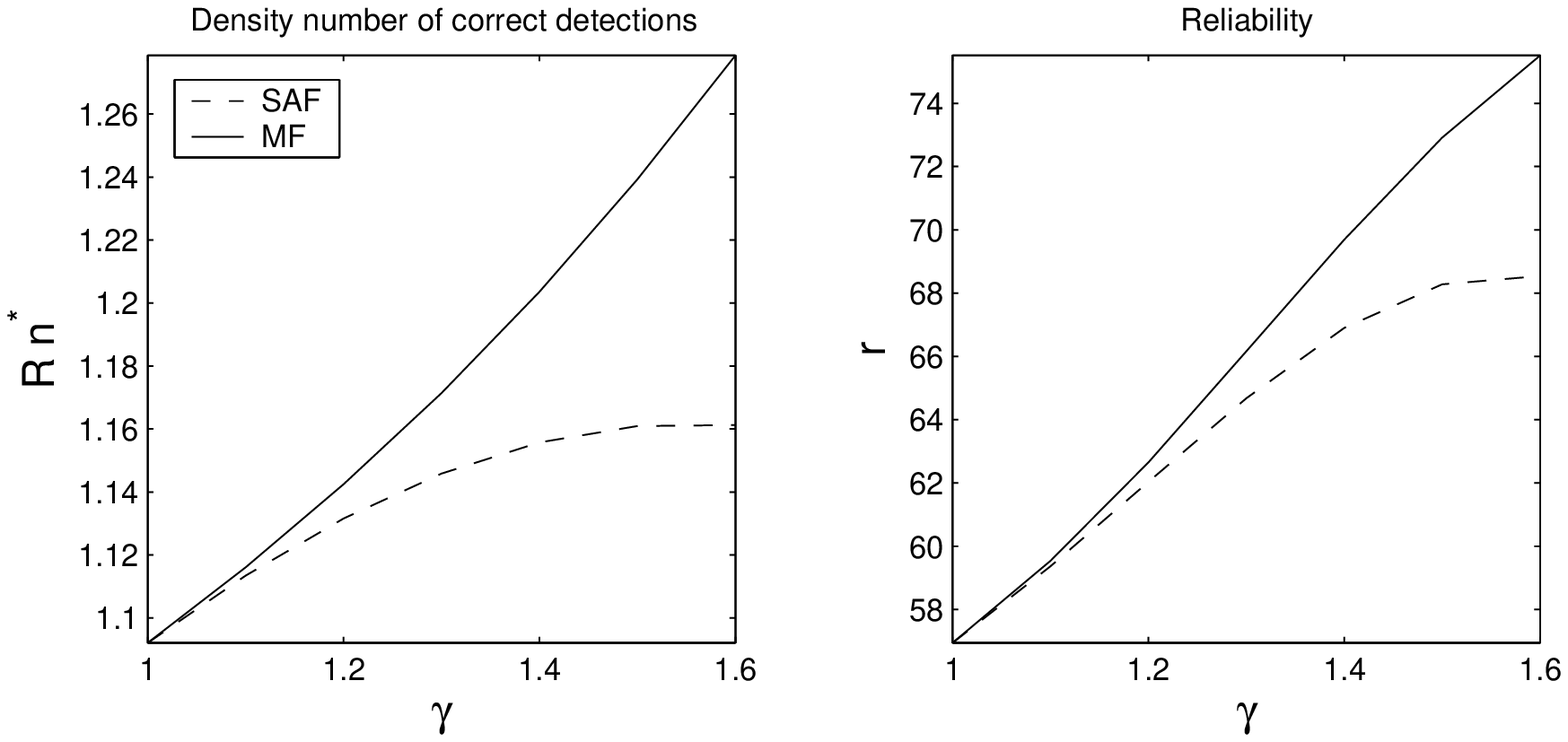}}
        \caption{Relationship between $\gamma$ vs. the
        number density $n^*$ of correct detections and the
        {\it reliability} $r$ corresponding to SAF and MF when for 
        both filters the number density of incorrect detections 
	  is fixed to the value $n_b^*$ of SAF with $L_* = 1$.}
        \label{fig:compar3}
\end{figure}

If one is worried of incorrect detections, there is a simple cure: the choice of
a $L_*$ making the detection of the sources less efficient.
In this way, part of the correct detections
will be lost but also the number of incorrect detections will decrease. Furthermore, in case of sources embedded
in the background the signal peaks are expected to have a mean height larger than that expected in case of only
background signal. Therefore, the smaller detection efficiency will affect more the number of incorrect detections 
than that
of the correct ones. This fact is shown in Fig.~\ref{fig:compar2} where it is evident that, when $\gamma=1.4$
and $L_* \approx 1.2$, MF has the same number density of incorrect detections as SAF with $L_*=1$ but still a larger 
number density of correct detections and consequently a larger
reliability $r$. The conclusion is that, as done in {\vtw}, a meaningful evaluation of the performances of the two 
filters requires that the comparison is made by fixing the number density of incorrect (or alternatively, correct)
detections.
If $n^*_b$ is set at the value of SAF for $L_* = 1$, the quantity $r$ shown in Fig.~\ref{fig:compar3} indicates that 
MF is better than SAF also for $1 \le \gamma \le 1.6$.

\begin{figure}
        \resizebox{\hsize}{!}{\includegraphics{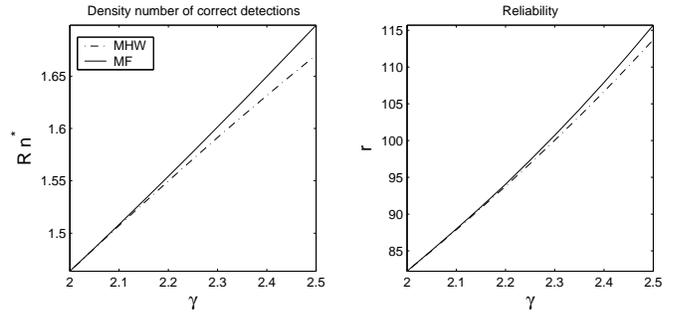}}
        \caption{Relationship $\gamma$ vs. the
        number density $n^*$ of correct detections and the
        {\it reliability} $r$ corresponding to MHW and MF when for 
        both filters the number density of incorrect detections 
	  is fixed to the value $n_b^*$ of MHW with $L_* = 1$.}
        \label{fig:compar3a}
\end{figure}

Similar arguments hold also for MHW that {\bshm} claim to provide a slightly better performance than MF 
when $\gamma \ge 2$. Fig.~\ref{fig:compar3a} shows again that this conclusion is not correct.

\subsection{Second case: random source amplitudes}

The arguments presented in the previous section have been developed under the hypothesis that all the sources
are characterized by the same amplitude $A$. Of course, this condition is not satisfied in real situations.
In order to solve this problem, {\bshm} suggest to substitute the likelihood ratio (\ref{eq:rule})
with:
\begin{equation}
L(\nu, \kappa) \equiv \frac{p(\nu, \kappa)}{p(\nu,\kappa | 0)},
\end{equation}
where
\begin{equation}
p(\nu, \kappa) = \int p(\nu,\kappa| \nu_s) p(\nu_s) d\nu_s,
\end{equation}
and $p(\nu_s)$ is the probability density function of the normalized amplitudes. After that, the method
of estimating the quantities $n_b^*$, $n^*$, $r$, and $D$ proceedes in the same way as in the previous 
section with the only exception that $n(\nu,\kappa|\nu_s)$ has to be substituted with $ n(\nu, \kappa) \equiv
\int n(\nu, \kappa | \nu_s) p(\nu_s) d\nu_s$. By assuming that $p(\nu_s)$ is an uniform probability distribution 
function with values in the 
range $[0 ,\nu_c]$, {\bshm} arrive to results similar to those they obtained for the case with fixed source amplitude. 
Again, this deserves some comments.

The first, and most obvious, is that such a conclusion suffers the same limitation
found in the previous section. Consequently the claim of superiority of SAF and MHW over MF is again not founded.
The second comment is that, in order to obtain reliable results, $p(\nu_s)$ is needed to
be known with good accuracy: the use of a wrong $p(\nu_s)$ will end in a false rule according to which
$p(\nu_s)$ overweighs the smallest amplitudes or the largest ones, favouring
the (correct and incorrect) detections or the (correct and incorrect) rejections with obvious consequences
on the ``optimality'' of the method.
In the framework of CMB studies the a priori information on $p(\nu_s)$ is not
available or is very inaccurate. The consequence is that a simple detection rule as, for example, the $3\sigma$ 
thresholding criterion
could still represent the best choice since it requires the only a priori knowledge of the noise level. This
approach is 
much simpler and safer than estimating the distribution of the source amplitudes.


\section{Summary and Conclusions} \label{sec:conclusions}

This paper deals with the detection techniques to extract point-sources from Cosmic Microwave Background maps.
Various recent works appeared in the literature, presenting new techniques 
with the aim to improve the performances of the classical matched filters (MF).
In particular the scale adaptive filters (SAF) and the Mexican hat wavelet (MHW) have been proposed as the most
efficient and reliable methods \citep[see][ and references therein]{san01}. This claim was subject to criticism
by \citet{vio02} since they showed that in reality SAF and MHW have performances that in general are
inferior to those provided by MF.

Recently \citet{bar03} used the argument that
a criterion making use of a simple $n\sigma$ thresholding rule is not fully sufficient to claim
detection. To support this assertion \citet{bar03}, in the context of one-dimensional signals and sources
with Gaussian profiles, adopt a 
detection criterion based on a Neyman-Pearson decision rule that makes use of both the height and the curvature 
of the maxima in the signal. Their theoretical arguments and numerical simulations indicate that, although
in general MF still remains the filter with the best performances, there are situations where SAF and
MHW overperform it.
In this paper we show that this conclusion is again not correct since
it is basically founded on a performance test favouring the filters characterized by a low 
detection capability. This means that there is no reason to prefer SAF or MHW to MF. Furthermore, the claimed superiority
of SAF and MHW, when the source scale has to be estimated from the data, has still to be
proved, and in principle also MF could be modified in such a way to efficiently deal with this situation.

These conclusions are not academic: the use of non-standard statistical tools 
is indicated only in situations
of real and sensible improvements of the results. New techniques that do not fulfill this requirement should
be introduced with care: they prevent the comparison with the results obtained in other works and may lead
people to use not well tested methodologies (MF has been successfully used for many years in very
different scientific contextes) ending up in not reliable results. Moreover, in the
present context, the use of SAF introduces further complications in the
analytical form of the filters (e.g., compare Eq.~(\ref{eq:filter2}) with Eq.~(\ref{eq:psi1})) and in the
definition of the detection rule (for MF the calculation of the curvature $\kappa$ of the peaks in the signal 
is not required).

\end{document}